\documentstyle[12pt,epsf]{article}

\begin{document}

\title{\Large \bf Predictions of $pp,\bar pp$ total cross section and\\
$\rho$ ratio at LHC and cosmic-ray energies\\
based on duality\footnote{Invited talk given at ``New Trends in High-Energy Physics"
    held at Yalta, Crimea(Ukraine), September 10-17, 2005.}}
\author{\large Keiji Igi$^1$, Muneyuki Ishida$^2$ \bigskip \\
{\it  $^1$~Theoretical Physics Laboratory, RIKEN, Wako, Saitama 351-0198, Japan} \\ 
{\it $^2$~Department of Physics, Meisei University, Hino, Tokyo 191-8506, Japan} \\
 (Presented by Keiji Igi)   }

\date{  }

\maketitle

{\large

\begin{center}
{\bf Abstract}\\
\end{center}
\medskip
Based on duality, we previously proposed to use rich informations on $\pi p$ total 
cross sections below $N (\sim 10$ GeV) in addition to high-energy data in order to 
discriminate whether these cross sections increase like log $\nu$ or log$^2\nu$ 
at high energies. We then arrived at the conclusion that our analysis prefers the
log$^2\nu$ behaviours. Using the FESR as a constraint for high energy 
parameters also for the $pp$, $\bar pp$ scattering,
we search for the simultaneous best fit to the data points of 
$\sigma_{\rm tot}$ and $\rho$ ratio up to some energy (e.g., ISR, Tevatron)
to determine the high-energy parameters. We then predict $\sigma_{\rm tot}$ and 
$\rho$ in the LHC and high-energy cosmic-ray regions. Using the data up to 
$\sqrt s=1.8$ TeV (Tevatron), we predict $\sigma_{\rm tot}^{pp}$ and $\rho^{pp}$ 
at the LHC energy ($\sqrt s=14$TeV)
as $106.3\pm 5.1_{\rm syst} \pm 2.4_{\rm stat}$mb and 
$0.126\pm 0.007_{\rm syst}\pm 0.004_{\rm stat}$, respectively.
The predicted values of $\sigma_{\rm tot}$ in terms of the same parameters
are in good agreement with the cosmic-ray experimental data up to 
$P_{lab}$ $\sim$ $10^{8\sim 9}$GeV.


\section{Introduction} \label{s1}
As you all know, the sum of $\pi^- p,\pi^+ p$ total cross sections 
has a tendency to increase above 70 GeV.
It had not been known before 2002, however, if this increase behaved like
log $\nu$ or log$^2\nu$ consistent with the Froissart-Martin bound\cite{[6]}.  
So, we proposed\cite{[1]} to use rich informations of $\pi p$ total cross sections 
at low energies in addition to high energy data in order to discriminate between 
asymptotic log $\nu$ or log$^2\nu$ behaviours, 
using a kind of the finite-energy sum sule (FESR) as constraints. 
Thus, duality is always satisfied in this approach.

Such a kind of attempt to investigate high-energy behaviours from those at 
low and intermediate energies has been initiated by one of the authors\cite{[3]}. 
In the early days of the Regge pole theory, there were controversies if 
there are other singularities with the vacuum quantum numbers 
except for the Pomeron (P). 
Under the assumption that no J singularities extend above $\alpha =0$ except for the Pomeron, 
we were led to the exact sum rule \cite{[3]} for the s-wave $\pi N$ scattering length $a^{(+)}$ 
of the crossing-even amplitude as
\begin{eqnarray}
\left( 1+\frac{\mu}{M} \right) a^{(+)} &=& -\frac{f^2}{M} + \int_0^N dk 
\left[ \sigma_{\rm tot}^{(+)}(k)-\sigma_{\rm tot}^{(+)}(\infty )  \right]  
 - \frac{\beta N^\alpha}{\alpha}\ \ .
\label{c1}
\end{eqnarray}
The evidence that Eq.~(\ref{c1}) was not satisfied empirically led to 
the $P^\prime$ trajectory with $\alpha_{P^\prime}\approx 0.5$ and the $f$ meson with spin two 
was discovered on the $P^\prime$ trajectory.

After 40 years, we have attempted\cite{[1]} to investigate whether the $\pi p$ total cross sections
increase like log $\nu$ or log$^2\nu$ at high energy based on the similar approach. 
We then arrived at the conclusion that our analysis prefers the log$^2\nu$ behaviours 
consistent with the Froissart-Martin unitarity bound. 
Recently, Block and Halzen\cite{[8],[a]} also reached the same conclusions 
based on duality arguments\cite{[4],[5]}.

\section{General approach}
Let us come to the main topics and begin by explaining how 
to predict $\sigma_{\rm tot}^{(+)}$, the $\bar pp$, $pp$ total cross 
sections and $\rho^{(+)}$, the ratio of the real to imaginary part of the forward scattering amplitude 
at the LHC and the higher-energy cosmic-ray regions, using the experimental data for 
$\sigma_{\rm tot}^{(+)}$ and $\rho^{(+)}$ for 70GeV$<P_{lab}<P_{large}$ as inputs. 
We first choose $P_{large}=2100$GeV corresponding to ISR region($\sqrt{s}\simeq 60$GeV). 
Secondly we choose $P_{large}=2\times 10^6$GeV corresponding to the Tevatron collider 
($\sqrt{s}\simeq 2$TeV). 
Let us search for the simultaneous best fit of $\sigma_{\rm tot}^{(+)}$ and $\rho^{(+)}$ 
in terms of high-energy parameters $c_0,c_1,c_2$ and $\beta_{P^\prime}$ 
constrained by the FESR. 
It turns out that the prediction of $\sigma_{\rm tot}^{(+)}$ agrees with $pp$ experimental data 
at these cosmic-ray energy regions\cite{Cosmic,[7]} within errors in the first case ( ISR ). 
It has to be noted that the energy range of predicted $\sigma_{\rm tot}^{(+)}$, $\rho^{(+)}$ 
is several orders of magnitude larger than the energy region of 
$\sigma_{\rm tot}^{(+)}$, $\rho^{(+)}$ input (see Fig.~\ref{fig:1}).  
If we use data up to Tevatron (the second case), 
the situation is much improved, although there are some systematic uncertainties
coming from the data at $\sqrt s=1.8$TeV (see Fig.~\ref{fig:2}).

\subsection{FESR(1)}
Firstly let us derive the FESR in the spirit of the $P^\prime$  
sum rule \cite{[3]}. Let us consider the crossing-even forward scattering amplitude defined by
\begin{eqnarray}
F^{(+)}(\nu ) &=& \frac{f^{\bar pp}(\nu )+f^{pp}(\nu )}{2}\ \    
{\rm with}\ \  Im\ F^{(+)}(\nu )=\frac{k\ \sigma^{(+)}_{\rm tot}(\nu )}{4\pi}\ .
\label{eq1}
\end{eqnarray}

We also assume 
\begin{eqnarray}
Im\ F^{(+)}(\nu ) &=& Im\ R(\nu )+ Im\ F_{P^\prime}(\nu )\nonumber\\
 &=& \frac{\nu}{M^2}\left( c_0 + c_1 {\rm log}\frac{\nu }{M} + c_2 {\rm log}^2\frac{\nu }{M}  \right)
  + \frac{\beta_{P^\prime}}{M}\left( \frac{\nu}{M} \right)^{\alpha_{P^\prime}}\ \ \ \ \ 
\label{eq2}
\end{eqnarray}
at high energies ($\nu > N$).  We have defined the functions $R(\nu )$ and $F_{P^\prime} (\nu )$ 
by replacing $\mu$ by M in Eq.~(3) of ref.\cite{[1]}.
Here, $M$ is the proton( anti-proton) mass and $\nu ,\ k$ are the incident proton(anti-proton) 
energy, momentum in the laboratory system, respectively.

Since the amplitude is crossing-even, we have
\begin{eqnarray}
R(\nu ) &=& \frac{i\nu}{2M^2}\left\{ 2c_0+c_2\pi^2 
  + c_1 \left({\rm log}\frac{e^{-i\pi}\nu }{M}+{\rm log}\frac{\nu}{M}\right) \right. \nonumber\\
&& \left.  + c_2 \left({\rm log}^2\frac{e^{-i\pi}\nu }{M} + {\rm log}^2\frac{\nu}{M}\right)  \right\}\ ,\ \ \ \ \ \ \ 
 \label{eq3}\\
F_{P^\prime}(\nu ) &=& -\frac{\beta_{P^\prime}}{M}
 \left( \frac{(e^{-i\pi}\nu /M)^{\alpha_{P^\prime}}
       +(\nu /M)^{\alpha_{P^\prime}}}{{\rm sin}\pi\alpha_{P^\prime}} \right),
\label{eq4}
\end{eqnarray}
and subsequently obtain 
\begin{eqnarray}
Re\ R(\nu ) &=& \frac{\pi\nu}{2M^2}\left( 
  c_1 + 2 c_2 {\rm log}\frac{\nu}{M} \right)\ ,\ \ \  
 \label{eq5}\\
Re\ F_{P^\prime}(\nu ) &=& -\frac{\beta_{P^\prime}}{M}
 \left( \frac{\nu}{M}\right)^{0.5}\ ,\ \ \ 
\label{eq6}
\end{eqnarray}
substituting  $\alpha_{P^\prime} =\frac{1}{2}$ in Eq.~(\ref{eq4}). Let us define
\begin{eqnarray}
\tilde F^{(+)}(\nu ) &=& F^{(+)}(\nu )-R(\nu )-F_{P^\prime}(\nu) \sim \nu^{\alpha (0)}
\ (\alpha (0)<0)\ .    
\label{eq7}
\end{eqnarray}
Using the similar technique to ref.\cite{[1]}, we obtain
\begin{eqnarray}
Re\ \tilde F^{(+)}(M) &=& \frac{2 P}{\pi} \int_0^\infty 
         \frac{\nu Im\ \tilde F^{(+)}(\nu )}{k^2} d\nu \ \nonumber\\    
 &=& \frac{2 P}{\pi} \int_0^M 
         \frac{\nu}{k^2} Im\ F^{(+)}(\nu ) d\nu 
    +\frac{1}{2\pi^2} \int_0^{\overline{N}} 
         \sigma_{\rm tot}^{(+)}(k) dk \ \nonumber\\    
 & & - \frac{2 P}{\pi} \int_0^N 
         \frac{\nu}{k^2} \left\{ Im\ R(\nu ) 
  + \frac{\beta_{P^\prime}}{M}\left(\frac{\nu}{M}\right)^{0.5}  \right\} d\nu \ , 
\label{eq8}
\end{eqnarray}
where $\overline{N}=\sqrt{N^2-M^2} \simeq N$.  Let us call Eq.~(\ref{eq8}) as the FESR(1).
If $c_1,c_2\rightarrow 0$, this Eq.~(\ref{eq8}) reduces to the so-called $P^\prime$ FESR 
in 1962\cite{[3]}.

\subsection{FESR(2)}
The second FESR corresponding to $n=1$ \cite{[5]} is:
\begin{eqnarray}  &&
\int_0^M \nu Im\ F^{(+)}(\nu )d\nu 
     + \frac{1}{4\pi}\int_0^{\overline{N}} k^2 \sigma_{\rm tot}^{(+)}(k)dk \nonumber\\
 & &=  \int_0^N \nu Im\ R(\nu ) d\nu 
     + \int_0^N \nu Im\ F_{P^\prime}(\nu ) d\nu \ \ \ . \ \ \  
\label{eq9}
\end{eqnarray}
We call Eq.~(\ref{eq9}) as the FESR(2) which we use in our analysis.

\subsection{The  $\rho^{(+)}$ ratio}
Let us obtain the $\rho^{(+)}$ ratio, 
the ratio of the real to imaginary part of $F^{(+)}(\nu )$, 
from Eqs.~(\ref{eq2}), (\ref{eq5}) and (\ref{eq6}) as
\begin{eqnarray}
\rho^{(+)}(\nu ) &=& \frac{Re\ F^{(+)}(\nu )}{Im\ F^{(+)}(\nu )}
  = \frac{Re\ R(\nu )+Re\ F_{P^\prime}(\nu )}{Im\ R(\nu )+Im\ F_{P^\prime}(\nu )} \nonumber\\
  &=& \frac{ \frac{\pi\nu}{2M^2}\left( c_1+2c_2 {\rm log} \frac{\nu}{M} \right) 
          -\frac{\beta_{P^\prime}}{M}\left(\frac{\nu}{M}\right)^{0.5} }{
                  \frac{k\sigma_{\rm tot}^{(+)}(\nu)}{4\pi} }\ .\ \ \ 
\label{eq10}
\end{eqnarray}

\subsection{General procedures} 
The FESR(1)(Eq.~(\ref{eq8})) has some problem. i.e., there are the so-called 
unphysical regions coming from boson poles below the $\bar pp$ threshold.
So, the contributions from unphysical regions of the first term of the right-hand side
of Eq.~(\ref{eq8}) have to be calculated.
Reliable estimates, however, are difficult. 
Therefore, we will not adopt the FESR(1).

On the other hand, contributions from the unphysical regions to the first term of the
left-hand side of FESR(2)(Eq.~(\ref{eq9})) can be estimated to be an order of
0.1\% compared 
with the second term.\footnote{The average of the imaginary part from boson
resonances below the $\bar pp$ threshold is the smooth extrapolation of the $t$-channel
$qq\bar q\bar q$ exchange contributions from high energy to $\nu\le M$ 
due to FESR duality\cite{[4],[5]}.
Since
$Im\ F^{(+)}_{qq\bar q\bar q}(\nu ) < Im\ F^{(+)}(\nu )$,
$\int_0^M \nu Im\ F^{(+)}_{qq\bar q\bar q}(\nu )d\nu 
< \int_0^M \nu Im\ F^{(+)}(\nu ) d\nu = \int_0^M \frac{\nu}{2} Im\ f^{\bar pp}(\nu ) d\nu
\simeq \frac{M^2}{4} \left. Im\ f^{\bar pp}\right|_{k=0}
\simeq 3.2{\rm GeV} \ll \frac{1}{4\pi} \int_0^{\overline{N}} k^2 \sigma_{\rm tot}^{(+)}(k)dk
=3403\pm 20$GeV, where we use the experimental value, 
$\frac{k}{4\pi}\sigma_{\rm tot}^{\bar pp}\simeq$14.4GeV$^{-1}$ in $k<\ 0.3$GeV.
So, resonance contributions to the first term of Eq.~(\ref{eq9}) is less than 0.1\% of 
the second term.

Besides boson resonances, there may be additional contributions from multi-pion contributions 
below $\bar pp$ threshold. In the $\bar pp$ annihilation, $\bar pp\rightarrow \pi\pi$ could 
give comparable contributions with $\rho$-meson, but multi-pion contributions are suppressed 
due to the phase volume effects. Therefore, the first term of Eq.~(\ref{eq9}) will still be 
negligible even if the above contributions are included.} 
Thus, it can easily be neglected.

Therefore, the FESR(2)(Eq.~(\ref{eq9})), 
 the formula of $\sigma_{\rm tot}^{(+)}$(Eqs.~(\ref{eq1}) and (\ref{eq2})) 
 and the $\rho^{(+)}$ ratio (Eq.~(\ref{eq10})) are our starting points.
Armed with the FESR(2), we express high-energy parameters
$c_0,c_1,c_2,\beta_{P^\prime}$ in terms of the integral of total cross sections up to
$N$. 
Using this FESR(2) as a constraint for $\beta_{P^\prime}=\beta_{P^\prime}(c_0,c_1,c_2)$,
the number of independent parameters is three.
We then search for the simultaneous best fit to the data points of $\sigma_{\rm tot}^{(+)}(k)$
and $\rho^{(+)}(k)$ for 70GeV$\le k \le P_{large}$ to determine the values of
$c_0,c_1,c_2$ giving the least $\chi^2$. 
We thus predict the $\sigma_{\rm tot}$ and $\rho^{(+)}$ 
in LHC energy and high-energy cosmic-ray regions.

\subsection{Data}
We use rich data\cite{[7]} of $\sigma^{\bar pp}$ and $\sigma^{pp}$ to evaluate the relevant 
integrals of cross sections appearing in FESR(2). 
We connect the each data point.
We then have  
\begin{eqnarray}
\frac{1}{4\pi} \int_0^{\overline{N}} k^2 \sigma_{\rm tot}^{(+)}(k) dk 
 & = & 3403\pm20\ {\rm GeV}.  
\label{eq11}
\end{eqnarray}
for $\overline{N}=10$GeV (which corresponds 
to $\sqrt s=E_{cm}=4.54$GeV).
%
(For more detail about data, see ref.\cite{[PLB]}.)

It is necessary to pay special attention to treat the data with the maximum 
$k=1.7266\times 10^6$GeV($\sqrt s=1.8$TeV) in this energy range,
which comes from the three experiments E710\cite{12d}$/$E811\cite{12c} and CDF\cite{12b}.
The former two experiments are mutually consistent and their averaged $\bar pp$ cross section
is $\sigma_{\rm tot}^{\bar pp}=72.0\pm 1.7$mb, which deviates from the result of 
CDF experiment $\sigma_{\rm tot}^{\bar pp}=80.03\pm 2.24$mb.

The two points of $\rho^{\bar pp}$ are reported in the SPS and Tevatron-collider energy region,
$1\times 10^5{\rm GeV}\le k \le 2\times 10^6$GeV (
at $k=1.5597\times 10^5$GeV($\sqrt s=541$GeV)\cite{rho1} and 
$k=1.7266\times 10^6$GeV($\sqrt s=$1.8TeV)\cite{12d} ).
We regard these two points as the $\rho^{(+)}$ data. 
As a result, we obtain 
9 points of $\rho^{(+)}$ up to Tevatron-collider energy region, 
$70{\rm GeV}\le k \le 2\times 10^6$GeV.

In the actual analyses, 
we use $Re\ F^{(+)}$ instead of $\rho^{(+)}(=Re\ F^{(+)}/Im\ F^{(+)})$.
The data points of $Re\ F^{(+)}(k)$ are made by multiplying $\rho^{(+)}(k)$ by
$Im\ F^{(+)}(k)=\frac{k}{8\pi}(\sigma_{\rm tot}^{\bar pp}(k)+\sigma_{\rm tot}^{pp}(k))$.

\subsection{Analysis} 
As was explained in the general procedure, 
both $\sigma_{\rm tot}^{(+)}$ and $Re\ F^{(+)}$ data in 70GeV $\le k \le P_{large}$ 
are fitted simultaneously through the formula Eq.~(\ref{eq2}) and Eq.~(\ref{eq10}) 
with the FESR(2)(Eq.~(\ref{eq9})) as a constraint. 
FESR(2) with Eq.~(\ref{eq11}) gives us
\begin{eqnarray}
8.87 &=& c_0 + 2.04 c_1 + 4.26 c_2 + 0.367 \beta_{P^\prime}\ ,
\label{eq12}
\end{eqnarray}
which is used as a constraint of $\beta_{P^\prime}=\beta_{P^\prime}(c_0,c_1,c_2)$,
and the fitting is done by three parameters $c_0,c_1$ and $c_2$.

We have done for the following three cases:\\
{\bf fit 1)}:\ \ \  The fit to the data up to ISR energy region,
       70GeV $\le k \le$ 2100GeV, 
       which includes 12 points of $\sigma_{\rm tot}^{(+)}$ 
       and 7 points of $\rho^{(+)}$. \\
{\bf fit 2)}:\ \ \  The fit to the data up to 
     Tevatron-collider energy region, 70GeV$\le k \le 2\times 10^6$GeV.
     For $k=1.7266\times 10^6$GeV($\sqrt s=1.8$TeV), the E710$/$E811 datum is used.
     There are 18 points of $\sigma_{\rm tot}^{(+)}$ 
       and 9 points of $\rho^{(+)}$.\\
{\bf fit 3)}:\ \ \  The same as fit 2, except for the CDF value at $\sqrt s=1.8$TeV, are used.

\subsection{Results of the fit} 
The results are shown in Fig.~\ref{fig:1}(Fig.~\ref{fig:2}) for the fit 1(fit 2 and fit 3).
The $\chi^2/d.o.f$ are given in Table \ref{tab1}. 
%
%
The reduced $\chi^2$ and the respective $\chi^2$-values devided by the number of data points 
for $\sigma_{\rm tot}^{(+)}$ and $\rho^{(+)}$ are less than or equal to unity.  
The fits are successful in all cases.
There are some systematic differences between fit 2 and fit 3, which come from the
experimental uncertainty of the data at $\sqrt s=1.8$TeV mentioned above.

\begin{table}
\caption{ The values of $\chi^2$ for the fit 1 (fit up to ISR energy) and 
         the fit 2 and fit 3 (fits up to Tevatron-collider energy). 
$N_F$ and $N_\sigma (N_\rho )$ are the degree of freedom and 
the number of $\sigma^{(+)}_{\rm tot}(\rho^{(+)})$ data points in the fitted energy region.
}
\begin{tabular}{c|ccc|}
         & $\chi^2/N_F$        
            &  $\chi^2_{\sigma}/N_{\sigma}$ & $\chi^2_{\rho}/N_{\rho}$ 
  \\
\hline
fit 1  & 10.6/15 & 3.6/12 & 7.0/7\\  
fit 2  & 16.5/23 & 8.1/18 & 8.4/9\\
fit 3  & 15.9/23 & 9.0/18 & 6.9/9\\
\hline
\end{tabular}
\label{tab1}
\end{table}

\begin{figure}
  \epsfxsize=16cm
  \epsfysize=20cm
 \centerline{\epsffile{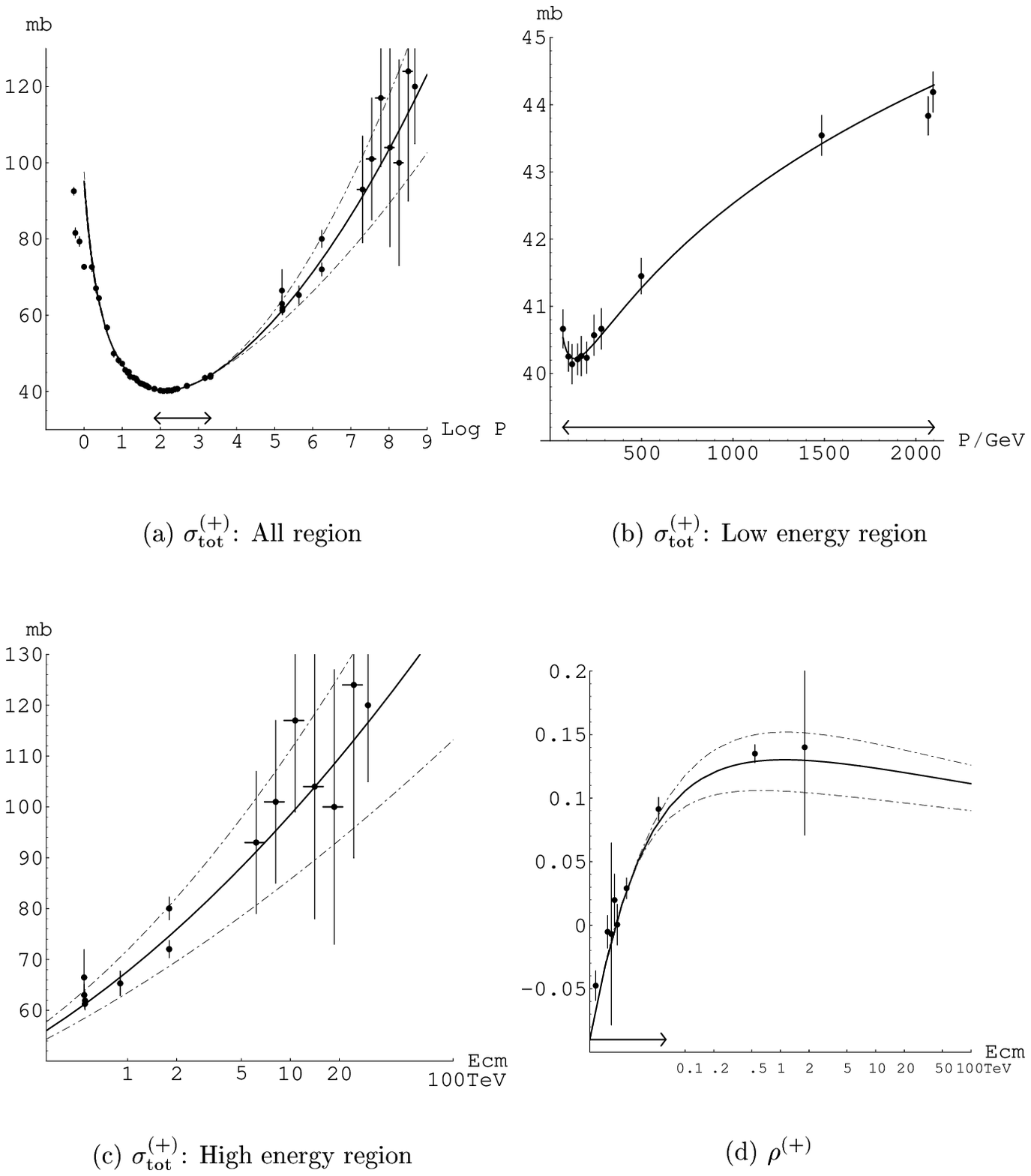}}
\caption{\label{fig:1} Predictions for $\sigma^{(+)}$ and $\rho^{(+)}$ in terms of the fit 1.
The fit is done for the data up to the ISR energy, in the region 70GeV$\le$ $k$ $\le$ 2100GeV
(11.5GeV $\le \sqrt s \le$ 62.7GeV) which is shown by the arrow in each figure. 
Total cross section $\sigma^{(+)}_{\rm tot}$ in 
(a) all energy region, versus log$_{10}P_{lab}/$GeV,
(b) low energy region (up to ISR energy), versus $P_{lab}/$GeV and
(c) high energy (Tevatron-collider, LHC and cosmic-ray energy) region, 
    versus center of mass energy $E_{cm}$ in TeV unit.
(d) gives the $\rho^{(+)}(=Re\ F^{(+)}/Im\ F^{(+)})$ in high energy region, 
    versus $E_{cm}$ in terms of TeV. 
The thin dot-dashed lines represent the one standard deviation.  }
\end{figure}

\begin{figure}
  \epsfxsize=16cm
  \epsfysize=20cm
 \centerline{\epsffile{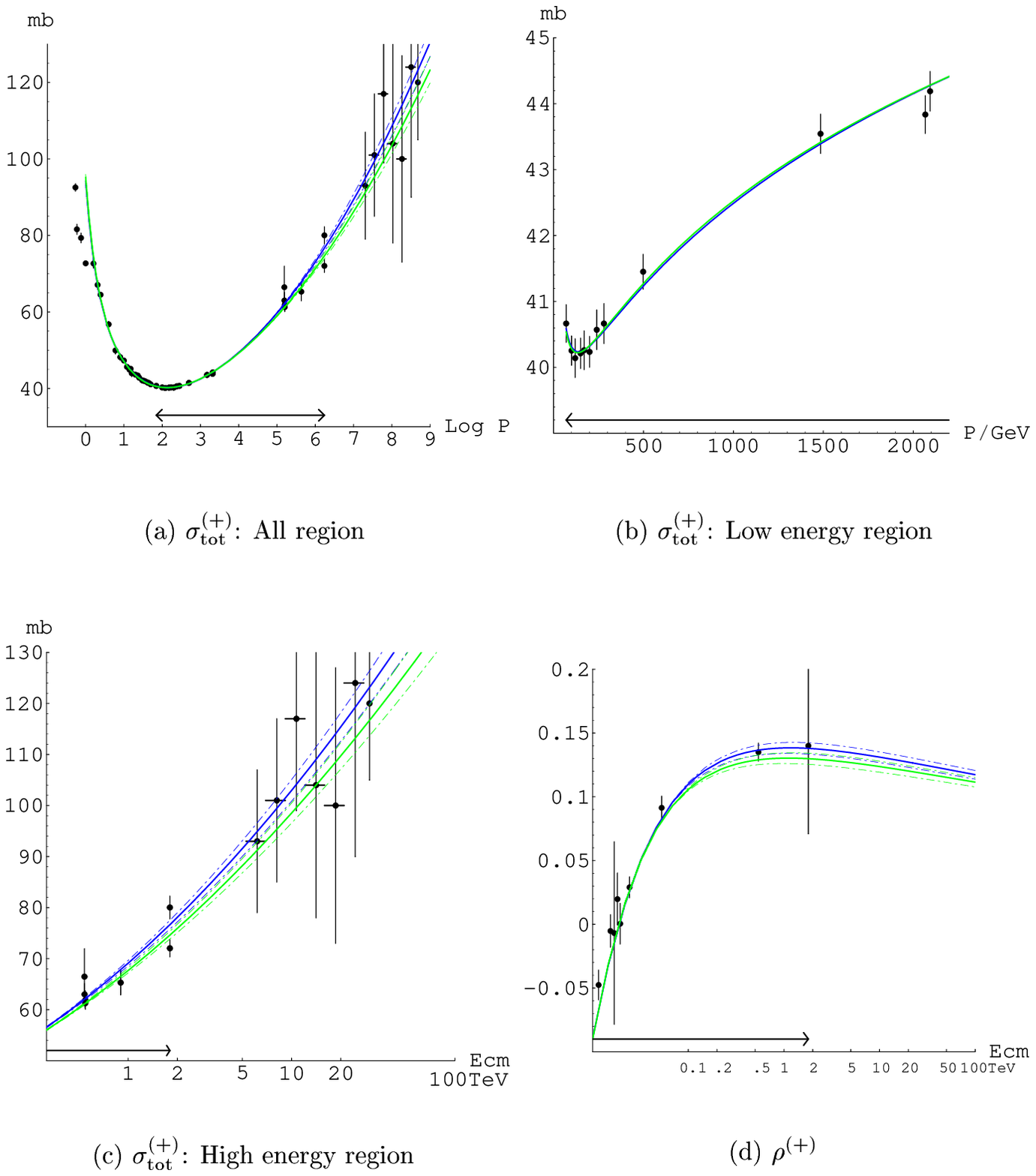}}
\caption{\label{fig:2} Predictions for $\sigma^{(+)}$ and $\rho^{(+)}$ in terms of 
the fit 2(shown by green lines) and fit 3(shown by blue lines).
The fit is done for the data up to Tevatron-collider energy, in the region  
70GeV$\le$ $k$ $\le$ $2\times10^6$GeV(11.5GeV $\le \sqrt s \le$ 1.8TeV) 
which is shown by the arrow. 
For $k=1.7266\times 10^6$GeV($\sqrt s=E_{\rm cm}=1.8$TeV), 
the averaged datum of E710\cite{12d}/E811\cite{12c}, $\sigma_{\rm tot}^{\bar pp}=72.0\pm 1.7$mb,
is used in fit 2, while the $\sigma_{\rm tot}^{\bar pp}=80.03\pm 2.24$mb of CDF\cite{12b} is used
in fit 3.
For each figure, see the caption 
in Fig.\ref{fig:1}.    }
\end{figure}

The best-fit values of the parameters are given in Table \ref{tab2}.
Here the errors of one standard deviation are also 
given.

\begin{table}
\caption{
The best-fit values of parameters in the fit 1, fit 2 and fit 3.
}
\begin{tabular}{c|cccc|}
       & $c_2$  & $c_1$  & $c_0$  &  $\beta_{P^\prime}$ \\
\hline
fit 1  & $0.0411\pm 0.0199$ & $-0.074\mp 0.287$ & $5.92\pm 1.07$ & $7.96\mp 1.55$ \\
fit 2  & $0.0412\pm 0.0041$ & $-0.076\mp 0.069$ & $5.93\pm 0.28$ & $7.95\mp 0.44$ \\
fit 3  & $0.0484\pm 0.0043$ & $-0.181\mp 0.071$ & $6.33\pm 0.29$ & $7.37\mp 0.45$ \\
\hline
\end{tabular}
\label{tab2}
\end{table}

\section{Predictions for $\sigma^{(+)}$ and $\rho^{(+)}$ 
at LHC and Cosmic-ray Energy Region}
 
By using the values of parameters in Table~\ref{tab2},
we can predict the $\sigma_{\rm tot}^{(+)}$ and $\rho^{(+)}$ in higher energy region, 
as are shown, respectively in (c) and (d) of Fig.~\ref{fig:1} and \ref{fig:2}. 
The thin dot-dashed lines represent the one standard deviation.

As is seen in (c) and (d) of Fig.~\ref{fig:1}, 
the fit 1 leads to
the prediction of $\sigma_{\rm tot}^{(+)}$ and $\rho^{(+)}$ with somewhat large errors in the 
Tevatron-collider energy region, although the best-fit curves are consistent 
with the present experimental data in this region. Furthermore, 
the predicted values of $\sigma_{\rm tot}^{(+)}$ agree with $pp$ experimental data 
at the cosmic-ray energy regions\cite{Cosmic,[e]} within errors (see (a),(c) of Fig.~\ref{fig:1}).
The best-fit curve gives $\chi^2/$(number of data) to be 13.0/16, 
and the prediction is successful.
As was mentioned before,
it has to be noted that the energy range of predicted $\sigma_{\rm tot}^{(+)}$
is several orders of magnitude larger than the energy region of 
the $\sigma_{\rm tot}^{(+)}$, $\rho^{(+)}$ input.  
If we use data up to Tevatron-collider energy region as in the fit 2 and fit 3, 
the situation is much improved (see (a),(c) of Fig.~\ref{fig:2}),
although there is systematic uncertainty depending on the treatment 
of the data at $\sqrt s=1.8$TeV.

The best-fit curve gives $\chi^2/$(number of data) from cosmic-ray data, 1.3/7(1.0/7)
for fit 2(fit 3).

We can predict the values of $\sigma_{\rm tot}^{(+)}$ and $\rho^{(+)}$
at LHC energy, $\sqrt s$=$E_{cm}$=14TeV and 
at very high energy of cosmic-ray region.
The relevant energies are very high, and
the $\sigma_{\rm tot}^{(+)}$ and $\rho^{(+)}$ can be regarded to be equal to the
$\sigma_{\rm tot}^{pp}$ and $\rho^{pp}$.
The results are shown in Table~\ref{tab3}.

\begin{table}
\caption{
The predictions of $\sigma_{\rm tot}^{(+)}$ and $\rho^{(+)}$ 
at LHC energy $\sqrt{s}=E_{cm}=14$TeV($P_{lab}$=1.04$\times 10^8$GeV), and 
at a very high energy $P_{lab}=5\cdot 10^{20}$eV
($\sqrt s$=$E_{cm}$=967TeV.) 
in cosmic-ray region.
}
\begin{tabular}{c|cc|cc|}
        &  $\sigma_{\rm tot}^{(+)}$({\scriptsize $\sqrt s$=14TeV}) 
        &  $\rho^{(+)}$({\scriptsize $\sqrt s$=14TeV})
        &  $\sigma_{\rm tot}^{(+)}$({\scriptsize $P_{lab}$=$5\cdot 10^{20}$eV}) 
        &  $\rho^{(+)}$({\scriptsize $P_{lab}$=$5\cdot 10^{20}$eV})\\
\hline
fit 1   & $103.8\pm 14.3$mb & $0.122\stackrel{+0.018}{\scriptstyle -0.024}$
        & $188\pm 43$mb & $0.099\stackrel{+0.011}{\scriptstyle -0.017}$\\
fit 2   & $103.8\pm 2.3$mb  & $0.122\pm 0.004$
        & $189\pm 8$mb  & $0.100\pm 0.003$\\
fit 3   & $108.9\pm 2.4$mb  & $0.129\pm 0.004$
        & $204\pm 8$mb  & $0.104\pm 0.003$\\
\hline
\end{tabular}
\label{tab3}
\end{table}

The prediction by the fit 1 
in which data up to the ISR energy are used as input has somewhat large(fairly large) errors 
at LHC energy(at high energy of cosmic ray). By including the data up to the Tevatron collider,
the prediction of fit 2(using E710/E811 datum) is smaller than that of fit 3(using CDF datum).
We regard the difference between the results of fit 2 and fit 3 as the systematic uncertainties
of our predictions. As a result,
we predict 
\begin{eqnarray}
\sigma_{\rm tot}^{pp} &=&  106.3\pm 5.1_{\rm syst} \pm 2.4_{\rm stat}\ {\rm mb},\ \   
\rho^{pp} = 0.126\pm 0.007_{\rm syst}\pm 0.004_{\rm stat}\ \ \ \ \ \ \ \ 
\label{eq13}
\end{eqnarray}
at LHC energy($\sqrt s=E_{cm}=14$TeV).
We obtain fairly large systematic errors coming from the experimental unceratinty 
at $\sqrt s=1.8$ TeV.


\section{Comparison with Other Groups}

The predicted central value of $\sigma_{\rm tot}^{pp}$ is in good agreement with 
Block and Halzen\cite{[a]} 
$\sigma_{\rm tot}^{pp}=107.4\pm 1.2$ mb, $\rho^{pp}=0.132\pm 0.001$. 
In contrary  to our results( see Fig.~2(a), (c)), however, their values are not affected 
so much about CDF, E710/E811 discrepancy. 
In our case, the measurements at LHC energy will discriminate which solution is better at 
Tevatron. 
Our prediction has also to be compared with 
Cudell et al.\cite{[f]} 
$\sigma_{\rm tot}^{pp}=111.5\pm 1.2_{\rm syst}\stackrel{+4.1}{\scriptstyle -2.1}_{\rm stat}$ mb, 
$\rho^{pp}=0.1361\pm 0.0015_{\rm syst}\stackrel{+0.0058}{\scriptstyle -0.0025}_{\rm stat}$, 
who's fitting techniques favour the CDF point at $\sqrt s=1.8$ TeV,
which leads to large value for $\sigma_{\rm tot}^{pp}$.
There are also predictions by Bourrely et al. \cite{[25]} $\sigma_{\rm tot}^{pp}=103.6$mb,
$\rho_{\rm tot}^{pp} = 0.122$, based on the impact-picture phenomenology.

%

Finally we emphasize that the LHC measurements would also clarify which
is the best solution among the three high-energy $\sigma^{pp}_{\rm tot}$ from $p$-air cross
sections\footnote{
The extraction of the $pp$ total cross section is based on the
determination of the proton-air production cross section from analysis of
extensive air shower. Detailed review \cite{[23]} on the subtleties involved are
found in refs.\cite{[c],[d],[e]}. The highest predictions for $\sigma^{pp}_{\rm tot}$ comes from
the results by Gaisser et al.\cite{[c]} and Nikolaev\cite{[d]}.  In the other extreme,
the lowest values come from the results by Block et al.\cite{[e]}. At the moment,
the predicted values of $\sigma_{\rm tot}^{pp}$ (see Fig.2) are in good agreement with
ref.\cite{[e]} since they are consistent with the Akeno results.

We would like to mention that it had already been pointed out by
Bourrely, Soffer and Wu \cite{[24]} that the Froissart bound is not merely an upper
bound but is actually saturated, i.e., the $\sigma_{\rm tot}^{pp}$ increases as log$^2$ $s$ 
for $s\rightarrow \infty$. There are also the phenomenological
predictions for higher energies in ref.\cite{[25]}. 
We were informed by S.F.Tuan about these works.}
\cite{[c],[d],[e]}.\\
{\it Acknowledgements}\ \ \ \ 
One of the authors (K.I.) would like to thank Prof. M.~Ninomiya and Prof. H.~Kawai
for their kind hospitality for completing this work, and also to Prof. L.~Jenkovszky
and the Organizing Committee for giving an opportunity to present this talk.
This work is supported by Grant-in-Aid for Scientific Research on
Priority Areas, Number of Area 763 ``Dynamics of Strings and Fields'',
from the Ministry of Education of Culture, Sports, Science and
Technology, Japan.

}


\begin{thebibliography}{99}
\bibitem{[6]} M.~Froissart, Phys.~Rev.~123 (1961) 1053.\\
 A.~Martin, Nuovo Cim. 42 (1966) 930.
\bibitem{[1]} K.~Igi and M.~Ishida, Phys.~Rev.~D 66 (2002) 034023.
\bibitem{[3]} K.~Igi, Phys.~Rev.~Lett.~9 (1962) 76.
\bibitem{[8]} M.~M.~Block and F.~Halzen, Phys.~Rev.~D 70 (2004) 091901.
\bibitem{[a]} M.~M.~Block and F.~Halzen, hep-ph/0506031.
\bibitem{[4]} K.~Igi and S.~Matsuda, Phys.~Rev.~Lett.~18 (1967) 625.
\bibitem{[5]} R.~Dolen, D.~Horn and C.~Schmid, 
Phys.~Rev.~166 (1968) 1768. 
This paper includes references on earlier papers on FESR.
\bibitem{Cosmic} M.~Honda et al. (Akeno Collab.), Phys.~Rev.~Lett. 70 (1993) 525.\\
R.~M.~Baltrusaitis et al. (Fly's Eye Collab.), Phys.~Rev.~Lett. 52 (1984) 1380.
\bibitem{[7]} Particle Data Group, S. Eidelman et al., 
Phys.~Lett.~B 592 (2004) 313.
\bibitem{ISR} G.~Carboni et al., Nucl.~Phys.~B 254 (1984) 697.\\
U.~Amaldi et al., Nucl.~Phys.~B 145 (1978) 367. 
\bibitem{TEVATRON} G.~Arnison et al., UA1 Collaboration, Phys.~Lett.~B 128 (1983) 336.\\
R.~Battiston et al., UA4 Collaboration, Phys.~Lett.~B 117 (1982) 126.\\
M.~Bozzo et al., UA4 Collaboration, Phys.~Lett.~B 147 (1984) 392.\\
G.~J.~Alner et al., UA5 Collaboration, Zeit.~Phys.~C 32 (1986) 153.
\bibitem{12a} C.~Augier et al., Phys.~Lett.~B 344 (1995) 451.
\bibitem{12d} N.~A.~Amos et al., E-710 Collaboration, Phys.~Rev.~Lett. 68 (1992) 2433.
\bibitem{12c} C.~Avila et al., E-811 Collaboration, Phys.~Lett.~B 445 (1999) 419.
\bibitem{12b} F.~Abe et al., CDF Collaboration, Phys.~Rev.~D 50 (1994) 5550.
\bibitem{rho} N.~Amos et al., Nucl.~Phys.~B 262 (1985) 689. 
\bibitem{rho1} C.~Augier et al., Phys.~Lett.~B 316 (1993) 448.
\bibitem{[PLB]} K.~Igi and M.~Ishida, Phys.~Lett.~B 622 (2005) 286.
\bibitem{[f]} J.~R.~Cudell et al., Phys.~Rev.~Lett.~89 (2002) 201801.
\bibitem{[23]} E.~G.~S.~Luna and M.~J.~Menon, Phys.~Lett.~B 565 (2003) 123.
\bibitem{[c]} T.~K.~Gaisser, U.~P.~Sukhatme and G.~B.~Yodh, Phys.~Rev.~D 36 (1987) 1350.
\bibitem{[d]} N.~N.~Nikolaev, Phys.~Rev.~D 48 (1993) R1904.
\bibitem{[e]} M.~M.~Block, F.~Halzen and T.~Stanev, Phys.~Rev.~D 62 (2000) 077501.
\bibitem{[24]} C.~Bourrely, J.~Soffer, T.~T.~Wu, Phys.~Rev.~D 19 (1979) 3249.
\bibitem{[25]} C.~Bourrely, J.~Soffer, T.~T.~Wu, Eur.~Phys.~J.~C 28 (2003) 97 and
references therein.

\end{thebibliography}
\end{document}